\documentclass{article}

\usepackage{microtype}
\usepackage{graphicx}
\usepackage{subfigure}
\usepackage{booktabs} % for professional tables

\usepackage{hyperref}

\usepackage{xcolor}

\usepackage[accepted]{mlsys2025}

\mlsystitlerunning{High-performance Resilient Serving}

\usepackage{xspace}
\newcommand{\sysname}{{FailSafe}\xspace~}

\begin{document}

\twocolumn[
\mlsystitle{FailSafe: High-performance Resilient Serving}

% unless you've provided the [accepted] option to the package.
\begin{mlsysauthorlist}
    \mlsysauthor{Ziyi Xu}{A}
    \mlsysauthor{Zhiqiang Xie}{B}
    \mlsysauthor{Swapnil Gandhi}{B}
    \mlsysauthor{Christos Kozyrakis}{B,C}
\end{mlsysauthorlist}

\mlsysaffiliation{A}{Shanghai Jiao Tong University}
\mlsysaffiliation{B}{Stanford University}
\mlsysaffiliation{C}{NVIDIA Research}

\mlsyscorrespondingauthor{Ziyi Xu}{xzy2022@sjtu.edu.cn}
\mlsyscorrespondingauthor{Christos Kozyrakis}{kozyraki@stanford.edu}

\mlsyskeywords{}

\vskip 0.3in

\begin{abstract}

Tensor parallelism (TP) enables large language models (LLMs) to scale inference efficiently across multiple GPUs, but its tight coupling makes systems fragile: a single GPU failure can halt execution, trigger costly KVCache recomputation, and introduce long-term compute and memory imbalance.
We present \sysname, a fault-tolerant TP serving system that sustains high performance under irregular GPU availability.
\sysname introduces three techniques to balance computation and memory across GPUs:
(1) Cyclic KVCache Placement for even memory utilization,
(2) Hybrid Attention combining tensor- and data-parallel attention to eliminate stragglers, and
(3) Fine-Grained Load-Aware Routing to dynamically balance requests.
It further employs proactive KVCache backup and on-demand weight recovery to avoid expensive recomputation and redundant data transfers.

We implement these techniques in a lightweight serving engine compatible with existing LLM infrastructures.
Evaluated on an 8×H100 DGX system with real-world fault traces and representative workloads, \sysname achieves up to 2× higher throughput and two orders of magnitude lower recovery latency compared to standard fault-handling approaches.
Even with up to three GPU failures, \sysname sustains high throughput and balanced utilization, demonstrating robust and efficient LLM serving under dynamic and unreliable hardware conditions.

\end{abstract}

] % this must go after the closing bracket ] following \twocolumn[ ...

\printAffiliationsAndNotice{Under Review}  % leave blank if no need to mention equal contribution

%%
%% Sections of Paper
%%
\section{Introduction}\label{sec:introduction}

Modern Large Language Models (LLMs) have significantly advanced capabilities across diverse applications, including conversational agents~\cite{chatgpt}, search engines~\cite{ggemini}, code generation~\cite{github-copilot}, and scientific discovery. As these models scale rapidly—now comprising hundreds of billions or even trillions of parameters—their computational and memory requirements have risen dramatically~\cite{deepseekv3}. Consequently, deploying LLM inference workloads increasingly demands aggregation of compute and memory resources across multiple GPUs, often distributed over several nodes. Tensor parallelism (TP) has emerged as a prominent solution for efficiently scaling these workloads by tightly coupling GPUs\footnote{We use ``GPU'' to refer to AI accelerators generally, such as GPUs, TPUs, and Trainium} using high-bandwidth interconnects such as NVIDIA’s NVLink~\cite{nvlink}, AMD’s infinity fabric link~\cite{amd-infinity} or Google's TPU Pods~\cite{google-tpu}.

While TP enables efficient scaling within a \emph{scale-up domain}—a set of GPUs connected via high-bandwidth interconnects—it also inherently couples all participating devices~\cite{megatron-lm}. This tight coupling creates a double-edged sword: a single GPU failure can render the entire TP execution unavailable across all GPUs within the affected scale-up domain. With scale-up domain sizes steadily increasing—with current deployments reaching up to 72 GPUs~\cite{NVL72} — the probability and impact of GPU failures both intensify. Recent studies~\cite{alibaba_fail, failure-characterization-in-datacenter, reliability-at-scale-meta, megascale} underline the rising frequency of GPU failures, attributing them to hardware degradation, thermal instability, and unpredictable preemptions. These failures impose substantial hurdles for maintaining reliable, high-performance inference services, critically affecting user experience and overall system efficiency. Typically, GPU failures introduce two primary types of overhead:

\vspace{0.5em}

\textbf{\textit{Challenge \#1: Recovery Overhead.}} When a GPU fails, several immediate complications arise in reassigning inference workloads. First, the KVCache for all inflight requests previously managed by the failed GPU is irrecoverably lost, requiring expensive and time-consuming recomputation before inference can continue. Additionally, the surviving GPUs must reshard and rebalance model weights (stored in CPU DRAM or persistent storage), resulting in substantial data movement and traffic over PCIe. Collectively, these recovery tasks—KVCache reconstruction and model weight resharing—trigger significant latency spikes, stall in-flight requests, and severely degrade the quality of experience for clients, as depicted in Figure~\ref{fig:breakdown_recovery_cdf}.

\vspace{0.5em}

\textbf{\textit{Challenge \#2:Persistent Imbalance Overhead.}} After recovery, the system continues with fewer GPUs than originally provisioned (e.g., from 8 to 7), breaking the symmetry assumptions under which serving pipelines are tuned. The result is enduring compute and memory skew. Compute imbalance occurs primarily due to uneven workload distribution, notably within attention layers partitioned by discrete attention heads. This imbalance leads to some GPUs idling, awaiting heavily loaded GPUs to complete their tasks, resulting in stalled resources and significantly diminished throughput. Concurrently, memory imbalance arises from unequal distribution of KVCache utilization across GPUs. This memory imbalance limits available cache capacity, forcing smaller batch sizes and further constraining throughput due to increased overhead from frequent kernel launches and reduced parallelism. These persistent imbalances, depicted in Figures~\ref{fig:design_memory_balance} and~\ref{fig:design_compute_balance}, consistently impose performance bottlenecks and cause severe underutilization of valuable GPU resources.

\vspace{1em}

To address these challenges, we propose \sysname, a system designed to provide high-performance yet resilient TP serving under irregular GPU availability.
\sysname introduces three key techniques to sustain efficiency and balance under irregular GPU availability:
(1) \textit{Cyclic KVCache Placement}, which rotates KVCache placement across layers to evenly distribute memory usage and preserve effective batch size;
(2) \textit{Hybrid Attention}, which integrates TP- and DP-style attention to balance computation across GPUs and eliminate stragglers; and
(3) \textit{Fine-Grained Load-Aware Routing}, which dynamically dispatches requests based on real-time GPU load to reduce intra-batch workload imbalance.

In addition, \sysname incorporates proactive KVCache backup to eliminate time-consuming recomputation during recovery, and on-demand weight restoration to minimize I/O overhead.
During normal operation, KVCache backups are asynchronously maintained in the background.
Upon failure, each GPU restores only the necessary subset of lost KVCache and model weights in a joint, non-redundant manner, thereby avoiding excessive PCIe data transfers and accelerating recovery.

We implement these mechanisms in a lightweight serving engine that integrates seamlessly with existing LLM infrastructures.
Evaluated on an 8$\times$H100 DGX system using real-world fault traces and representative serving workloads, \sysname demonstrates strong performance across both offline throughput–optimized and online latency–constrained scenarios.
It achieves up to $2\times$ higher throughput and up to two orders of magnitude lower recovery latency compared to standard fault-handling practices.
Further analysis shows that \sysname continues to deliver substantial performance benefits even when up to three of eight GPUs fail, highlighting its robustness and scalability under severe fault conditions.

\section{Resilient Model Serving}\label{sec:background}

\subsection{Background}

\subsubsection{LLM Inference}

Most popular LLMs are based on the Transformer architecture, comprising multiple stacked transformer layers~\cite{attention}. Each layer consists of an attention mechanism and a feed-forward network (FFN). Attention layers enable tokens within a request to interact, while FFN layers process tokens independently. Most LLMs adopt the auto-regressive decoding mechanism, leading to an incremental inference process consisting of several iterations~\cite{orca}. At each inference iteration, the model predicts the next token based on all previously generated tokens. To optimize performance and avoid redundant computation, LLM serving systems cache intermediate token states—known as the KVCache—for reuse in subsequent token generation steps~\cite{sglang, vllm}. This caching strategy divides the inference process into two distinct phases: the prefill phase, which processes all input tokens simultaneously in a single iteration to construct the initial key-value cache and generate the first output token; and the decoding phase, where only the newly generated token requires computation to update the KVCache.

\subsubsection{Multi-GPU LLM Inference}

Modern LLMs frequently exceed the memory capacity and performance capabilities of single GPUs, necessitating distributed execution strategies. Inference parallelism differs fundamentally from training parallelism, as it involves only forward-pass computations—without gradients—and introduces specific challenges such as efficient management of the key-value (KV) cache. Effective parallelism schemes typically aim either to minimize latency for individual inference requests or to maximize the overall throughput of concurrent workloads. Common parallelism strategies include data parallelism (DP), where the model is replicated across multiple nodes to independently handle separate requests, and pipeline parallelism (PP), where model layers are distributed sequentially across nodes to balance computational load and memory usage. Among these methods, tensor parallelism (TP)—which partitions model parameters and KVCache evenly across all GPUs within a node—stands out. TP uniquely leverages high-bandwidth interconnects within a node to reduce inference latency by splitting large matrix operations across GPUs. Hybrid Parallelism combines TP and PP, often using TP within compute nodes and PP across nodes, to scale to larger models and GPU counts.

\subsubsection{Frequent Failures in Large-Scale Clusters}\label{sub:avail}

Modern inference-serving clusters routinely consist of thousands of GPUs interconnected by sophisticated networking, storage, and power infrastructures. At this scale, hard failures—complete, sudden, and persistent hardware disruptions—occur frequently and inevitably. Common sources of these hard failures include GPU overheating, ECC errors, CUDA errors, and GPU driver errors leading to abrupt termination of system software running on them~\cite{failure-characterization-in-datacenter}. For instance, Alibaba Cloud \cite{alibaba_fail} reported abnormal termination rates as high as 44\% for its top 5\% most resource-intensive workloads. Similar reliability challenges have been echoed by Meta~\cite{reliability-at-scale-meta}, ByteDance~\cite{megascale}, LAION~\cite{laion}, and Google~\cite{tpu-resiliency}.

\subsection{Motivation and Challenges}
\label{sec:motivation}

\subsubsection{
Imbalanced Load over Irregular Hardware Configurations}

When deploying LLMs on an irregular number of GPUs, for example, seven instead of the ideal eight due to GPU availability issed discussed in \S\ref{sub:avail}, it becomes challenging to maintain balanced computation and memory usage across devices.
For feed-forward (FFN) layers, weights are typically partitioned along the intermediate dimension, allowing relatively even distribution even across fewer GPUs, since the intermediate dimension is large enough to divide smoothly.
In contrast, attention layers are divided by attention heads, which are usually limited to only tens per layer, leading to severe imbalance when sharded across an irregular number of GPUs.

For instance, when sharding a LLaMA-3.1-70B model \cite{llama} with 8 key-value heads across seven GPUs, some ranks inevitably host two heads while others hold only one, resulting in up to a $2\times$ slowdown in attention-layer computation (Figure~\ref{fig:design_compute_balance}).
This imbalance is further exacerbated in long-context scenarios, where each attention head contributes not only to computation but also to the KVCache memory footprint.
As a result, certain ranks may consume up to twice as much KVCache memory as others.
Due to the synchronized nature of tensor parallelism, such imbalance effectively reduces the usable batch size of the entire system, leading to substantial throughput degradation, as shown in Figure~\ref{fig:design_memory_balance}.

\subsubsection{Latency Spike due to State Loss}
\label{sub:recovery}

When a GPU suddenly fails or becomes temporarily unavailable, all model parameters and KVCache residing on it are immediately lost. While static model weights can be reloaded from persistent storage such as host memory or disk with moderate cost,  the KVCache represents the dynamic per-request state that must be recomputed from scratch. This recomputation requires rerunning the entire prefill phase for all affected in-flight requests, which is an extremely compute-intensive process. In our online serving experiments (Section~\ref{sec:recovery_latency}), recomputing the lost KVCache alone takes over 20 seconds, during which all affected requests undergo severe stalls.
Moreover, this backlog propagates through the serving pipeline, introducing queuing delays that further impact subsequent requests. The result is a sharp latency spike and widespread SLO violations, significantly degrading user-perceived quality of service.
This phenomenon highlights the urgent need for fast recovery mechanisms. Instead of recomputing from scratch, a more effective strategy is needed to speed up this entire recovery process.

\section{Design and Implementation}

To enable model execution on an irregular number of GPUs, we first implement a vanilla form of \textit{non-uniform tensor parallelism} for model serving, originally proposed in prior work on fault-tolerant LLM training~\cite{arfeen2025nonuniform}.
Non-uniform tensor parallelism allows model weights and computations to be distributed unevenly across GPUs by adjusting synchronization among tensor-parallel workers. This enables flexible tensor-parallel configurations under partial GPU availability.
To address the load imbalance and latency spikes discussed in \S\ref{sec:motivation}, we further introduce a memory and computation balancer and a lighting recovery mechanism for higher throughput and better SLO attainment.

\subsection{Memory and Computation Balancer}
\label{design:balancer}

\begin{figure}[ht]
    \centering
    \includegraphics[width=\linewidth]{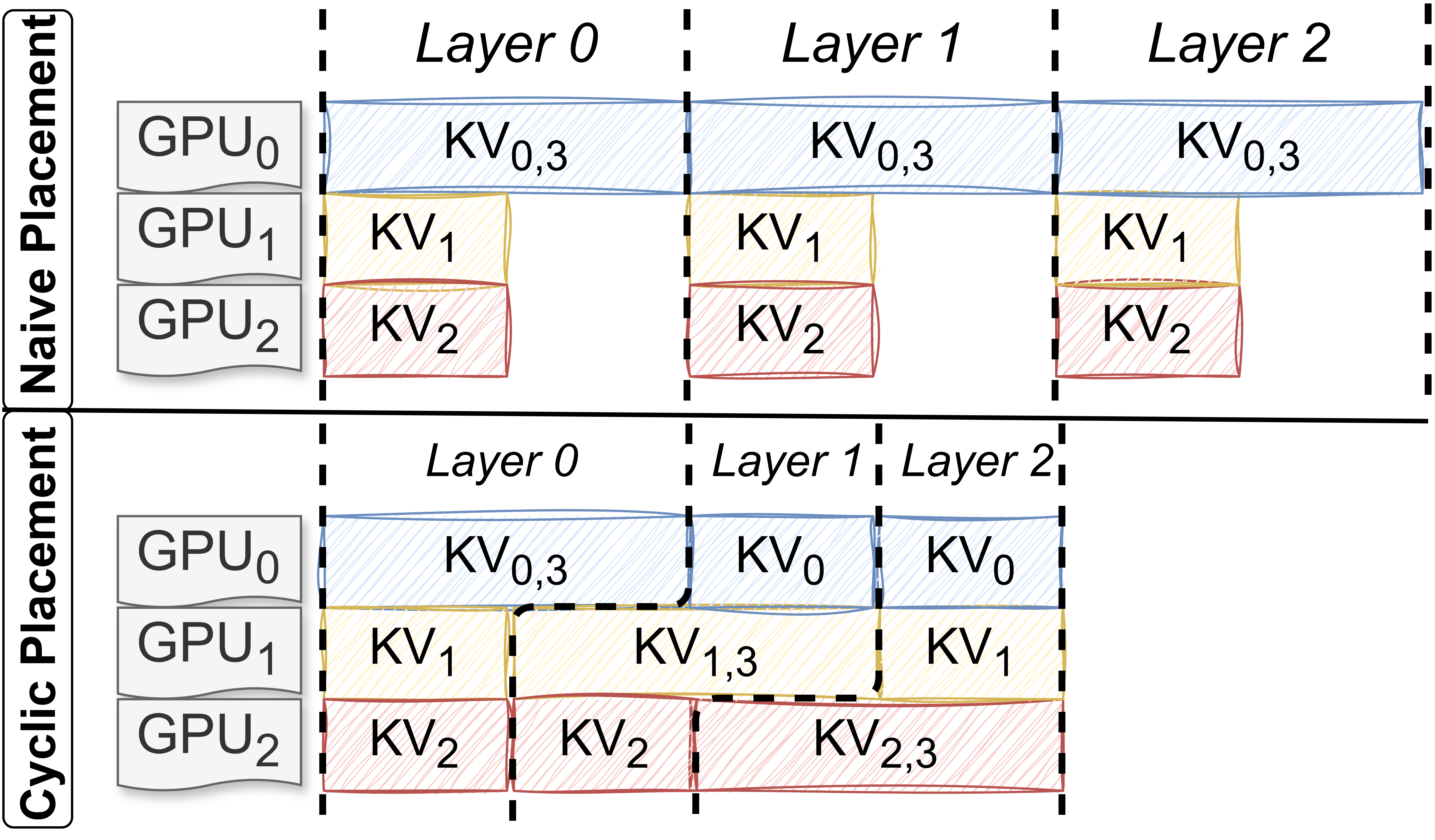}
    % \vspace{-2em}
    \caption{Illustration of the proposed cyclic placement for balancing KVCache memory usage across GPUs. In this example, the model has 4 key-value heads and deploys non-uniform TP3. KV$_{i}$ stands for the KVCache for $i$-th key value head. Cyclic placement (bottom) improves overall KVCache capacity by approximately 50\%, compared with a naïve placement (top).
    }
    \label{fig:design_memory_balance}
\end{figure}

\textbf{Cyclic Placement.} 
To mitigate memory imbalance, we propose a simple yet effective cyclic placement strategy. Specifically, attention heads and their corresponding KVCache blocks are distributed cyclically across GPUs.
As illustrated in Figure~\ref{fig:design_memory_balance}, naïve placement can lead to significant KVCache skew, where some GPUs accumulate substantially larger memory footprints. In contrast, our cyclic scheme rotates attention-head assignments layer by layer, ensuring that across every contiguous $n$ layers in a TP$n$ configuration, the aggregate KVCache allocation remains well balanced.
Because modern LLMs typically contain tens or even hundreds of layers—far exceeding the usual tensor-parallel world size (often below ten)—this cyclic distribution effectively smooths out KVCache memory utilization across all GPUs.

However, this strategy alone does not fully resolve computational imbalance. Each attention layer still performs an all-reduce operation before and after attention computation, synchronizing results across all GPUs. Thus, while cyclic placement balances memory consumption across layers, within each layer some GPUs continue to perform more computation than others, leading to temporary stalls.

\textbf{Hybrid Attention.}
To further mitigate intra-layer computation imbalance, we introduce Hybrid Attention, which maintains a balanced workload among tensor-parallel workers by assigning each the same number of attention heads and leveraging data parallelism (DP) to handle the remaining heads.
Hybrid Attention can be viewed as a generalization of TP and DP.

\begin{figure}[ht]
    \centering
    \includegraphics[width=\linewidth]{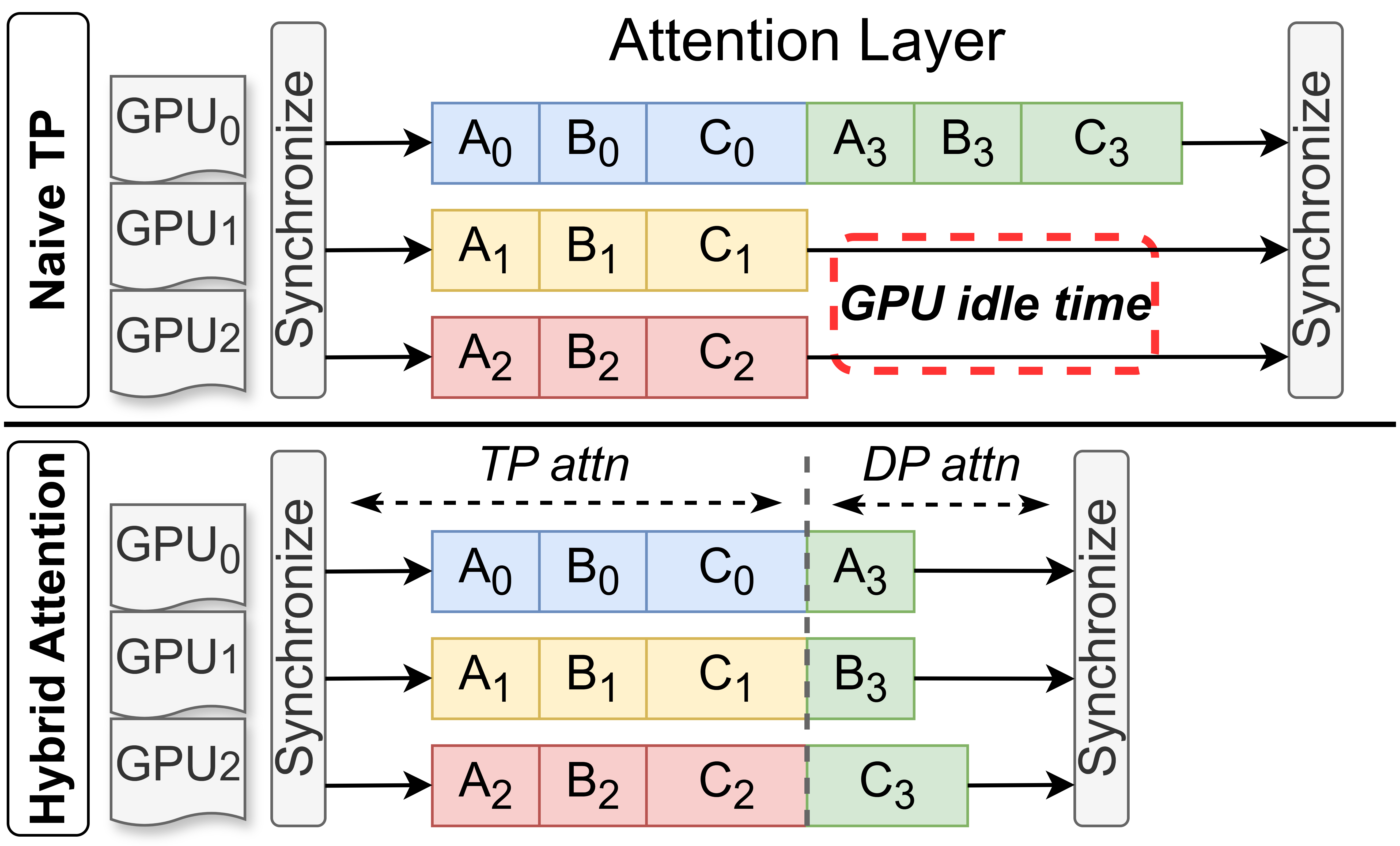}
    % \vspace{-2em}
    \caption{Illustration of the proposed hybrid attention. In this example, the model has 4 key-value heads and deploys non-uniform TP3. A$_i$ stands for the $i$-th head's computation for request $A$. Request $A$ is routed to GPU$_0$, $B$ to GPU$_1$, and $C$ to GPU$_2$. Hybrid attention (bottom) combine TP attention with DP attention, significantly reducing GPU idle time and improving GPU utilization.}
    \label{fig:design_compute_balance}
\end{figure}

For example, in the LLaMA-3 70B model with eight attention heads, a deployment with eight TP workers evenly distributes the heads, behaving identically to a standard TP-8 configuration.
However, when only seven GPUs—and hence seven TP workers—are available, each TP worker receives one head, while the remaining head is replicated across all seven GPUs. Each GPU thus also acts as a DP worker, processing the replicated head for different requests.
This design enables the computation of the extra head to be parallelized across GPUs through data parallelism, distributing requests to balance the workload effectively.
As shown in Figure~\ref{fig:design_compute_balance}, naïve non-uniform tensor parallelism often results in straggler GPUs during attention computation, causing under-utilization.
In contrast, Hybrid Attention distributes the computation of replicated heads for different requests across multiple GPUs, significantly reducing intra-layer imbalance, minimizing idle time, and improving overall utilization.

Notably, the widely adopted DP attention design~\cite{sglang2024v04}—used for serving multi-head latent attention (MLA) \cite{MLA} models such as DeepSeek-V3 \cite{deepseekv3}—is a special case of our Hybrid Attention.
While DP attention duplicates a single attention head across all GPUs, Hybrid Attention generalizes this approach to flexibly support both partitioned (TP) and replicated (DP) heads within the same layer.

However, DP alone does not guarantee perfect balance, since input requests can be inherently skewed. Even if all DP replicas process the same number of requests, some GPUs may experience longer execution times due to variable input lengths. In highly skewed workloads—especially in long-context scenarios—the system may temporarily revert to behavior similar to naïve non-uniform TP, where faster GPUs must still wait for the longest-running task.

\textbf{Fine-grained Load-aware Router.}
To further mitigate the load imbalance caused by skewed input request distributions, we introduce a fine-grained, load-aware routing mechanism that dynamically balances the workload among DP workers across GPUs.
Our scheduler design addresses two complementary aspects: (1) \textit{routing DP ranks}, and (2) \textit{forming compute-balanced batches}.

\textit{Load-Aware DP-Rank Routing.}
We observe that the DP-rank scheduling problem can be modeled as an instance of the classical \textit{online makespan minimization} problem~\cite{online_makespan_1,online_makespan_2,online_makespan_3}.
To achieve low scheduling overhead while maintaining balance, we adopt a simple yet effective greedy strategy: each incoming request is assigned to the GPU with the smallest estimated remaining workload.
Here, the workload is defined as the total pending DP computation (in token units) currently queued on each GPU.
This dynamic routing policy continuously adapts to runtime variations in request arrival patterns, preventing load concentration on specific GPUs.

\textit{Fine-Grained Chunked Prefill.}
To complement routing, we design a \textit{Fine-Grained Chunked Prefill} mechanism for the prefill stage.
Unlike conventional chunked prefill, which allows only one chunk per request in a batch. Our approach permits chunks from multiple requests to be executed jointly within the same batch.
Since the computational cost of prefill attention grows quadratically with sequence length, the cost of processing a chunk of size $N$ after $L$ processed tokens is $\mathcal{O}(N^2 + NL + N)$.
To maintain balanced GPU workloads and prevent out-of-memory errors, we define a global prefill token budget $N$ representing the maximum number of tokens per batch.
Tokens are then iteratively allocated to the least-loaded GPU until the budget is reached, ensuring balanced computation and bounded intermediate memory usage.
Algorithm~\ref{alg:dp_scheduler} outlines the adaptive chunked prefill procedure.

\begin{algorithm}[t]
\caption{DP-aware Adaptive Chunked Prefill}
\label{alg:dp_scheduler}
\begin{algorithmic}
    \STATE {\bf Input:} Token budget $N$, rank set $\mathcal{R}$, schedulable tokens $\{S_r\}$, workloads $\{W_r\}$
    \STATE {\bf Output:} Next Prefill batch $B$
    \STATE $L_r \!\leftarrow\! 0$ \COMMENT{initialize load per rank}
    \STATE $B \!\leftarrow\! \emptyset$ \COMMENT{initialize current prefill batch}
    \STATE $H\!\leftarrow\!\emptyset$ \COMMENT{initialize candidate batch set}
    \WHILE{$|B| < N$ \AND $\exists r:S_r\neq\emptyset$}
        \STATE $r^\ast \!\leftarrow\! \arg\min_{r \in \mathcal{R},\, S_r \neq \emptyset} L_r$ \COMMENT{least-loaded rank}
        \STATE $t \!\leftarrow\! \mathrm{first}(S_{r^\ast}),~~S_{r^\ast} \!\leftarrow\! S_{r^\ast}\!\setminus\!\{t\}$ \COMMENT{schedule 1 token}
        \STATE $B \!\leftarrow\! B\!\cup\!\{t\},~~L_{r^\ast} \!\leftarrow\! L_{r^\ast} + \mathrm{cost}(t), H\!\leftarrow\!H\cup\{B\}$
    \ENDWHILE
    \STATE \textbf{return} choose\_best\_batch($H$)
\end{algorithmic}
\end{algorithm}

\begin{figure}[ht]
    \centering
    \includegraphics[width=\linewidth]{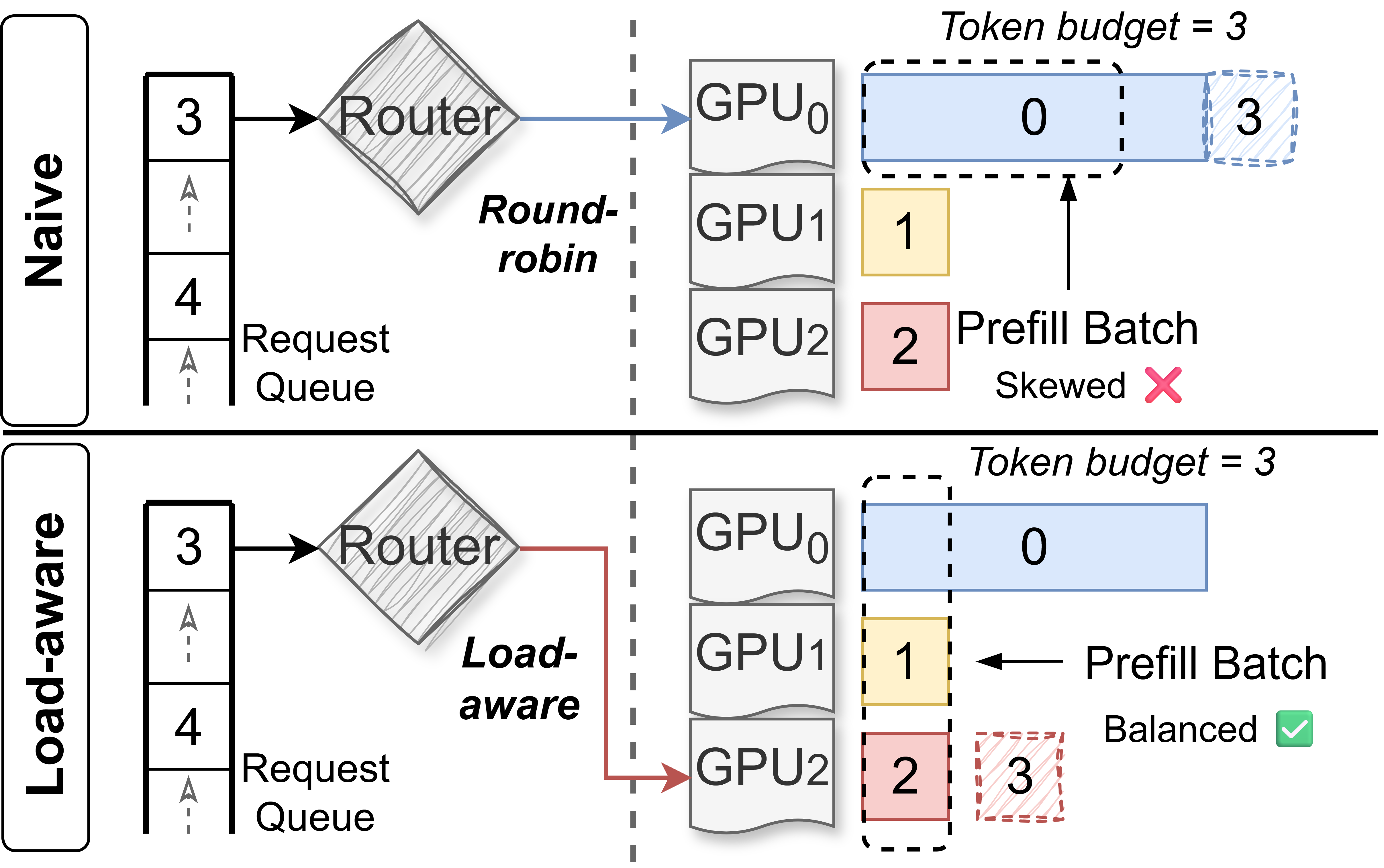}
    % \vspace{-2em}
    \caption{Illustration of the load-aware router and scheduler. In this example, request~0 has 4 tokens, request~1 and 2 has 1 token, and a new request~3 with 1 token arrives. In the naïve setting (top), a round-robin router combined with a FIFO chunked prefill scheduler results in an highly unbalanced prefill batch: within the prefill token budget (which is 3), only a chunk of request~0 is scheduled in the prefill batch. In contrast, our load-aware router (bottom) dynamically redirects new requests to the least-loaded GPU, and our adaptive chunked prefill mechanism helps form a balanced batch.
    }
    \label{fig:design_scheduler}
\end{figure}

As shown in Figure~\ref{fig:design_scheduler}, naïve routing and chunked prefill lead to highly skewed batches and unbalanced GPU workloads.
In the example, GPU$_0$ becomes overloaded because the vanilla scheduler includes only the first chunk of request~0 within the prefill budget, leaving other GPUs underutilized.
In contrast, our load-aware router dynamically assigns DP ranks based on real-time GPU load, while the adaptive chunked prefill scheduler constructs balanced batches in a best-effort manner.
Together, these two mechanisms significantly improve GPU utilization and overall system throughput under dynamic and skewed request patterns.

\subsection{Lightning Recovery}
As discussed in \S\ref{sub:recovery}, rapid recovery is critical to prevent request backlogs and maintain SLO compliance.
We identify two dominant sources of recovery latency: recomputation of the lost KVCache and reloading of re-sharded model weights.
To mitigate both, we propose a \textit{Lightning Recovery} framework comprising two components—\textit{Proactive KVCache Backup} and \textit{On-Demand Weight Recovery}—that together minimize downtime by avoiding redundant recomputation and restoring only essential model data.

\textbf{Proactive KVCache Backup.}
When a GPU fails, its associated KVCache is lost, requiring a costly re-prefill phase to regenerate it. To enable fast recovery, \sysname proactively backs up KVCache data to host memory. Modern GPU servers typically equip ample and persistent CPU memory that are larger than the GPU HBMs, which remains intact even after GPU faults, making it ideal for lightweight state preservation.
Restoring KVCache from host memory is often substantially faster than recomputing it as long as PCIe bandwidth is well utilized~\cite{xie2025strata}, offering both practicality and efficiency.

Upon failure, surviving GPUs reuse their existing KVCache, while each GPU reloads only its required portion from host memory. Thanks to the cyclic KV placement optimization discussed in \S\ref{design:balancer}, the reloaded cache is evenly distributed across GPUs, balancing PCIe transfer bandwidth during recovery.

\textbf{On-demand Weight Recovery.}
Reusing and restoring the KVCache eliminates recomputation overhead, but model weight reloading can still be a bottleneck.
To address this, \sysname leverages a key mathematical property: the sharding order of FFN weights can be freely permuted without affecting correctness, as matrix multiplication is commutative along the reduction dimension.

In conventional tensor-parallel inference, FFN weights are sharded contiguously along the intermediate dimension, so when the TP world size changes (e.g., from 8 to 7 due to a GPU fault), existing shards misalign with new ranks, forcing full-shard reloads.
By exploiting the commutativity of matrix multiplication, \sysname allows each rank to reload only the required subset of weights instead of an entire shard, significantly reducing data transfer and recovery time.
For attention layers, \sysname employs DP-based placement within its hybrid attention design.
To minimize redundant PCIe transfers, each rank loads a distinct portion of the DP weights from host memory and exchanges missing segments with peers via NVLink.
Because NVLink offers far higher bandwidth than PCIe, the synchronization overhead is minimal and can be overlapped with ongoing weight loading.
Together, these two strategies ensure that all ranks participate equally in recovery, fully utilizing NVLink and PCIe bandwidth while minimizing end-to-end recovery latency.

\begin{figure}[ht]
    \centering
    \includegraphics[width=\linewidth]{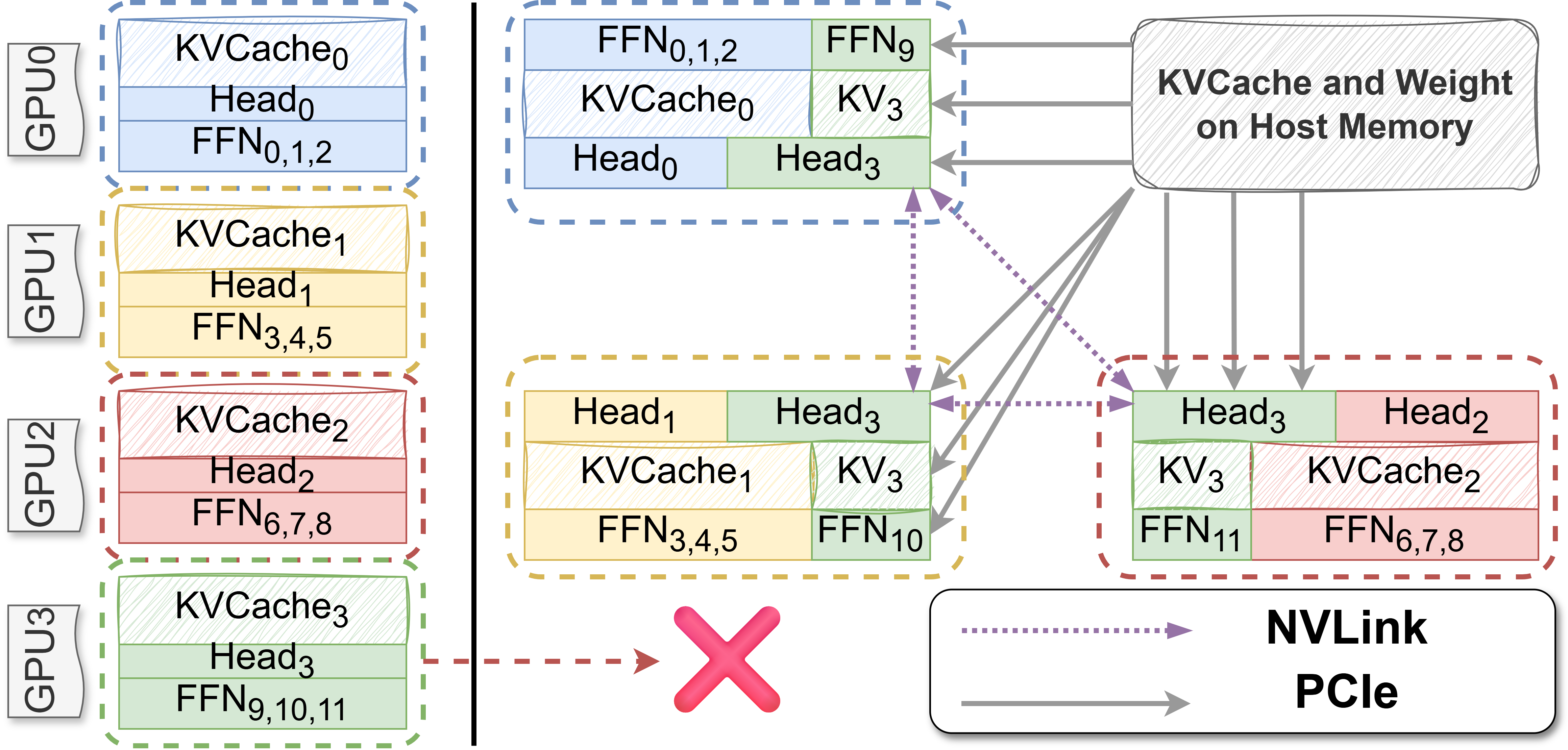}
    \caption{On-demand Recovery mechanism. In this example, the FFN weights are divided into 12 shards and there're 4 attention heads with corresponding KVCache. The system starts with normal TP4. When GPU 3 fails, we will restore all the lost state (weights and KVCache) via PCIe. Our On-demand recovery mechanism eliminate all redundant PCIe transfer.
    }
    \label{fig:recovery}
\end{figure}

Figure~\ref{fig:recovery} illustrates the on-demand recovery process under a TP4 setup. The model has 12 FFN shards, 4 attention heads, and 4 KVCache partitions distributed across three GPUs. When GPU3 fails, a naïve TP3 fallback would force GPU3 to load the entire missing shards (FFN$_{9,10,11}$) and attention head$_3$.
In contrast, \sysname enables each surviving GPU to reload only a fraction of the missing weights by leveraging FFN commutativity.
For the lost attention head, hybrid attention divides the required weights across GPUs, synchronizing the remaining segments via high-bandwidth NVLink.
Meanwhile, each rank recovers only the necessary subset of the missing KVCache from host memory, completing recovery quickly and in parallel.

\section{Evaluation}

We evaluate \sysname on a server equipped with eight NVIDIA H100 GPUs (80~GB each).
The GPUs are interconnected via 4th-generation NVLink and each GPU is connected to the CPU via a PCIe 5.0 x16 link.
We measure both system offline throughput as well online throughput latency characterization to demonstrate the overall effectiveness of our system.
To ensure a comprehensive evaluation, we experiment with two representative large language models that normally utilize a whole node of GPUs to serve: a dense model, LLaMA-3.1-70B-Instruct~\cite{llama}, and a mixture-of-experts (MoE) model, Mixtral-8x22B-Instruct-v0.1~\cite{mixtral}.
We implemented \sysname on top of a light-weight customized LLM serving engine, which is implemented using around 7k lines of code and achieves performance on par with state-of-the-art open-source LLM inference frameworks such as vLLM \cite{vllm} and SGLang \cite{sglang}.

\subsection{Offline Throughput under Faulty Environments}
\label{sec:offline_throughput}

\begin{figure}[h]
    \centering
    \includegraphics[width=\linewidth]{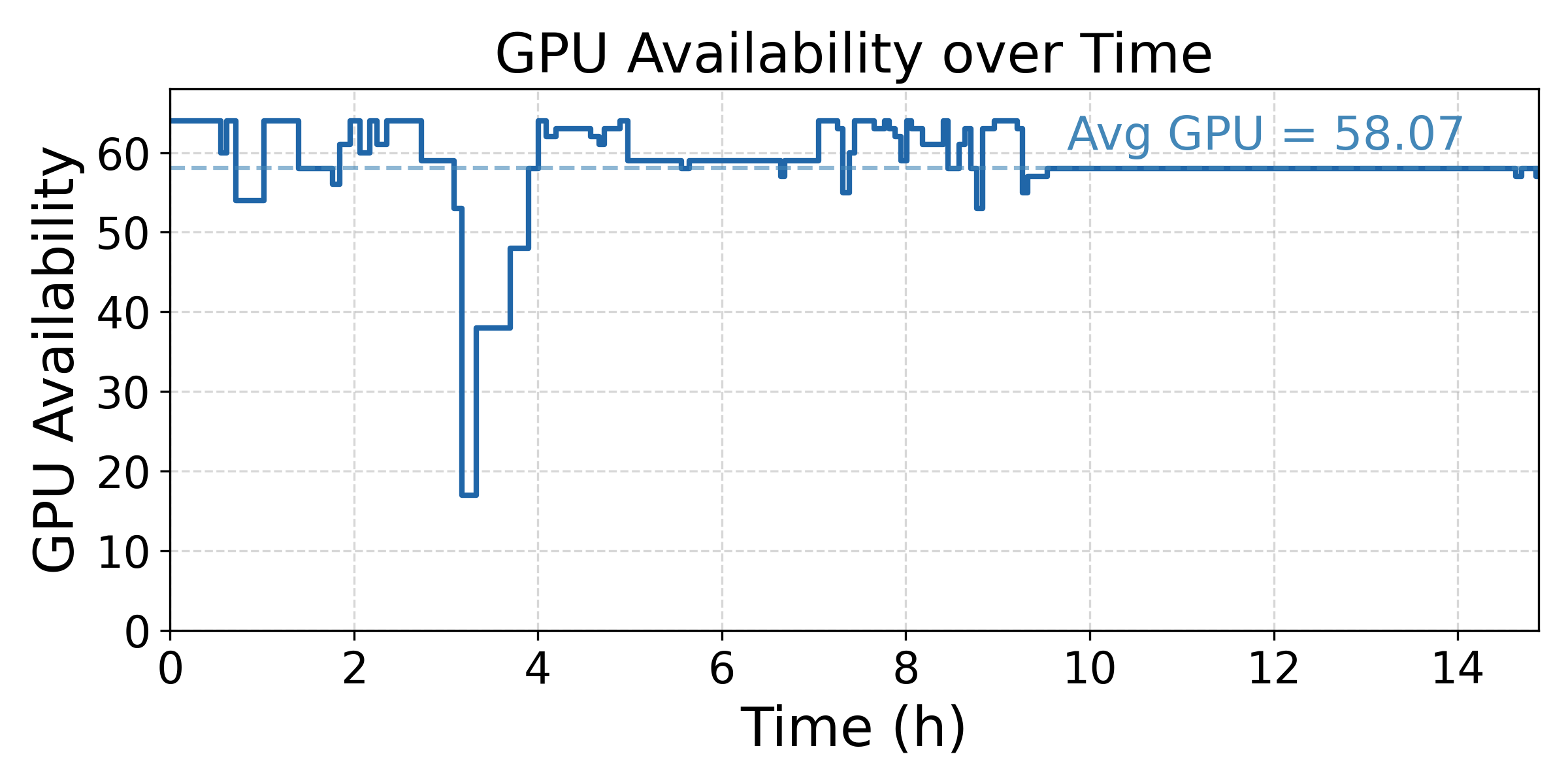}
    \caption{GPU availability from GCP cloud availability traces.}
    \label{fig:gcp_traces}
\end{figure}

\textbf{Failure Simulation.}
We simulate failure and recovery events on an $8 \times 8$ H100 cluster using a real-world failure trace derived from the GCP cloud availability dataset, which has been widely adopted in prior studies such as Bamboo~\cite{bamboo}, Oobleck~\cite{oobleck}, and Recycle~\cite{recycle}.
The trace is scaled such that full availability corresponds to 64 GPUs, as illustrated in Figure~\ref{fig:gcp_traces}.
During the simulation, each failure event randomly disables one GPU across the eight nodes, while each recovery event randomly restores one of the failed GPUs.
Upon every GPU failure, the system must reconfigure itself to continue operation with the reduced number of GPUs.
For simplicity, we fix the reconfiguration (switching) latency to 10 seconds for all systems, as this delay has negligible impact on the overall throughput.

\textbf{Workload Dataset.}

We use \textit{OpenThoughts-114k}~\cite{openthoughts}, a large-scale, high-quality “thinking” dataset composed of 114,000 multi-turn reasoning and instruction-following examples curated from diverse sources.
This dataset is widely used for evaluating and post-training LLMs on reflective and logical reasoning tasks, making it representative of realistic LLM-serving workloads.
Its key input–output characteristics are summarized in Table~\ref{tab:offline_openthought_stats}.
We adopt this dataset to emulate long-context, multi-turn interactions that commonly arise in RL-based training and inference workloads~\cite{rl}.

\begin{figure*}[t!]
\centering
% ---------- left: dense 70B ----------
\begin{minipage}{0.48\linewidth}
\centering
\begin{tabular}{c|ccccccccc}
\toprule
\textbf{Available GPUs} & 1 & 2 & 3 & 4 & 5 & 6 & 7 & 8 \\ 
\midrule
\textbf{Baseline System} & -- & -- & -- & 4 & 4 & 4 & 4 & 8 \\
\textbf{\sysname} & -- & -- & 3 & 4 & 5 & 6 & 7 & 8 \\
% \textbf{Oracle System} & 8 & 8 & 8 & 8 & 8 & 8 & 8 & 8 & 8 \\
\bottomrule
\end{tabular}
\vspace{0.3em}
\includegraphics[width=\linewidth]{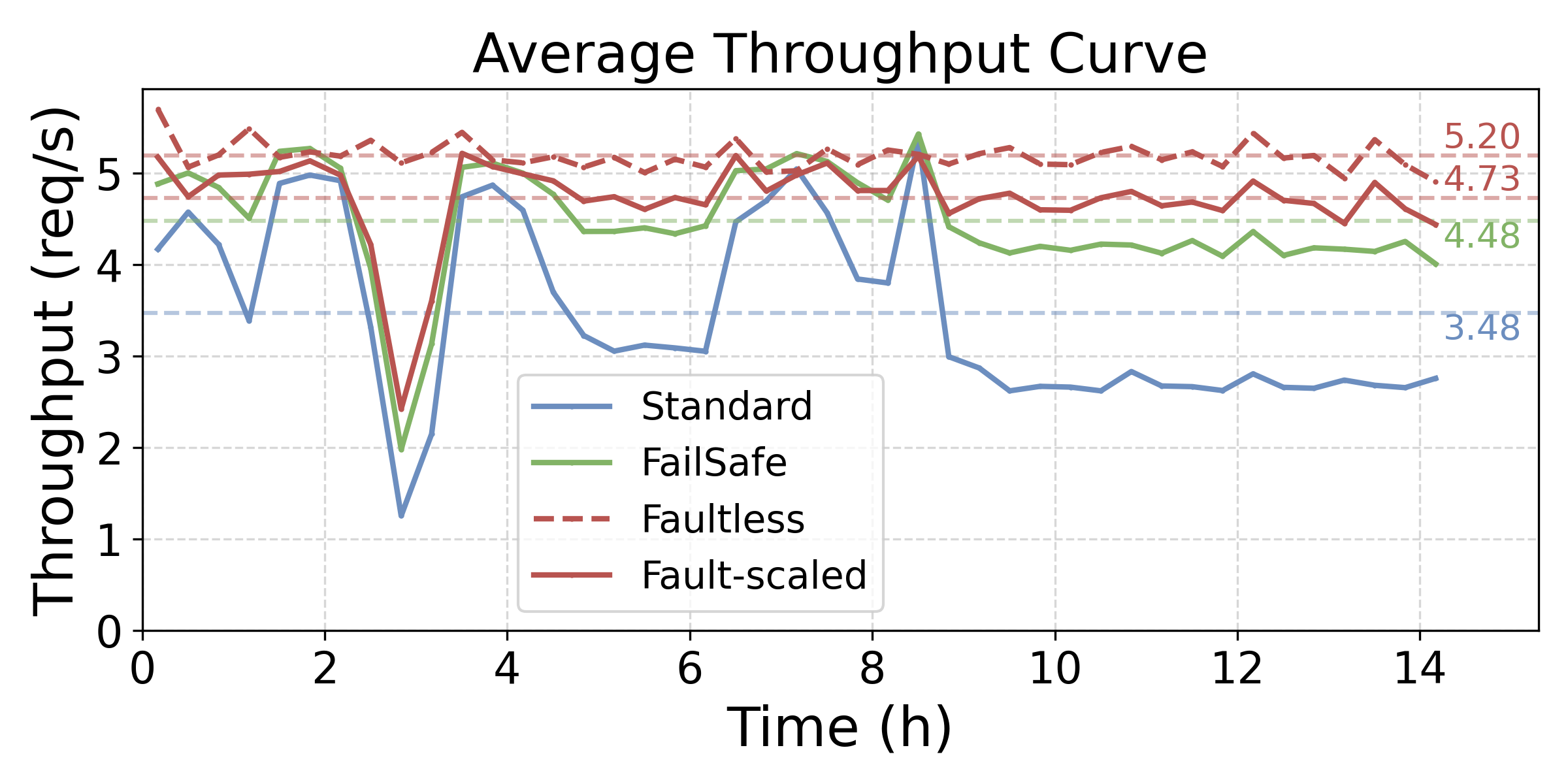}
\vspace{-1.5em}
\caption{(a) LLaMA-3.1-70B-Instruct}
\label{fig:offline_thought_dense}
\end{minipage}
\hfill
% ---------- right: sparse MoE ----------
\begin{minipage}{0.48\linewidth}
\centering
\begin{tabular}{c|ccccccccc}
\toprule
\textbf{Available GPUs} & 1 & 2 & 3 & 4 & 5 & 6 & 7 & 8 \\ 
\midrule
\textbf{Baseline System} & -- & -- & -- & -- & -- & -- & -- & 8 \\
\textbf{\sysname} & -- & -- & -- & -- & 5 & 6 & 7 & 8 \\
% \textbf{Oracle System} & 8 & 8 & 8 & 8 & 8 & 8 & 8 & 8 & 8 \\
\bottomrule
\end{tabular}
\vspace{0.3em}
\includegraphics[width=\linewidth]{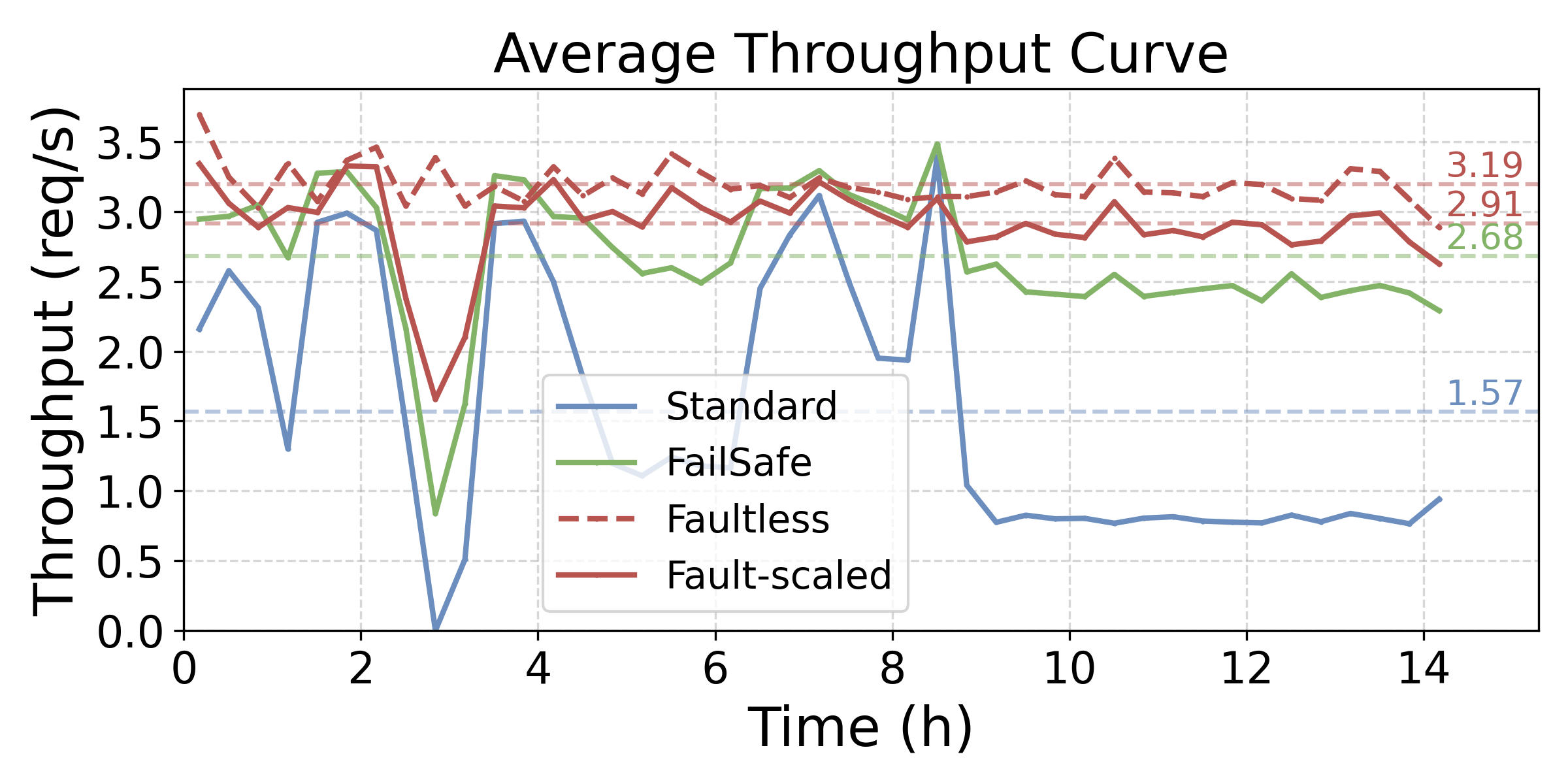}
\vspace{-1.5em}
\caption{(b) Mixtral-8x22B-Instruct-v0.1}
\label{fig:offline_thought_sparse}
\end{minipage}
\caption{
The tables above summarize the tensor-parallel configurations adopted by each system under different numbers of available GPUs per node. 
The figures below show the corresponding real-time throughput of different systems during fault-injection experiments. 
Dashed lines in each figure represent the average throughput over time.
}
\label{fig:offline_combined}
\end{figure*}

\begin{table}[h]
\begin{tabular}{lccc}
\toprule
\textbf{Metric} & \textbf{Mean} & \textbf{Median} & \textbf{Max} \\
\midrule
Input length (tokens)  & 422  & 352 & 7633 \\
Output length (tokens) & 7295  & 5583 & 37817 \\
\bottomrule
\end{tabular}
\centering
\caption{Input-output characteristics of OpenThought dataset.}
\label{tab:offline_openthought_stats}
\end{table}

\textbf{Baselines.}
We compare \sysname against three configurations:
(1) a non-fault-tolerant \textit{baseline} system,
(2) a \textit{fault-free} system that assumes no GPU failures and serves as the performance upper bound, and
(3) a \textit{fault-scale} system that linearly scales the throughput of the \textit{fault-free} case based on GPU availability.
We use a single 8-GPU machine to emulate eight independent 8-GPU nodes and report aggregated throughput across all simulated nodes.
All systems employ tensor parallelism within each node.
In the \textit{baseline} system, the per-node TP configurations are limited to ${1, 2, 4, 8}$ GPUs, consistent with the implementations of state-of-the-art serving engines such as SGLang and vLLM.
Consequently, when a GPU fails within a node, the \textit{baseline} system must fall back to the next supported configuration, resulting in reduced resource utilization.
In contrast, \sysname supports flexible TP configurations with arbitrary GPU counts, provided sufficient memory is available for model weights and KVCache.
The same flexibility applies to the \textit{fault-scale} setting.
As shown in Figure~\ref{fig:offline_combined}, the minimum number of GPUs required to serve the dense LLaMA-3.1-70B-Instruct model is three.
For the Mixture-of-Experts model Mixtral-8x22B-Instruct-v0.1, the larger memory footprint raises this minimum to five GPUs.

\begin{figure*}[t]
    \centering
    \includegraphics[width=\linewidth]{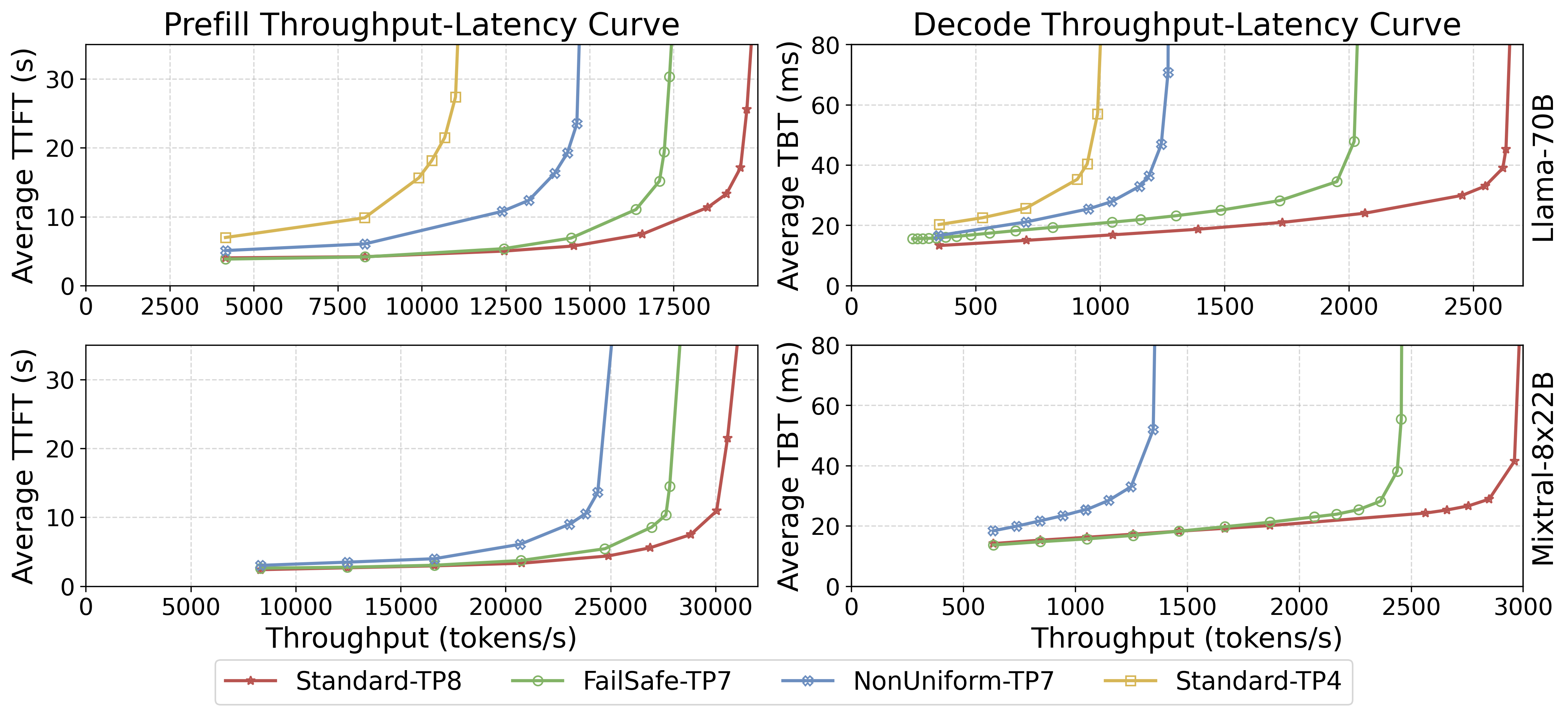}
    \vspace{-2em}
    \caption{
Throughput-latency curves when serving the Mooncake trace. 
The TP4 baseline is omitted for the Mixtral-8x22B model due to insufficient memory to accommodate both model weights and KVCache on only 4 GPUs.
    }
    \label{fig:e2e_online}
\end{figure*}

\textbf{Performance Analysis.}
As shown in Figure~\ref{fig:offline_combined}, \sysname consistently outperforms the standard TP4 baseline across both dense and MoE models.
For \textit{LLaMA-3.1-70B}, \sysname delivers 1.28$\times$ higher average throughput, achieving 95\% of the \textit{Fault-scaled} performance, demonstrating the effectiveness of its memory and compute balancing optimization.
For \textit{Mixtral-8x22B}, the throughput gain increases to 1.71$\times$, reaching 92\% of the \textit{Fault-scaled} performance.
The larger throughpout improvement stems from Mixtral’s higher memory footprint, where the standard TP4 configuration becomes infeasible, leading to greater resource underutilization that \sysname effectively mitigates.

\subsection{Throughput-Latency of Fault-tolerant Online Serving}
We further evaluate the throughput–latency characteristics of \sysname under online serving scenarios.
To ensure consistency, we measure throughput and latency by varying the input request rate in stable environments, i.e., the system operates with a fixed number of available GPUs and no runtime reconfiguration.
We separately analyze the \textit{prefill} and \textit{decode} stages, as prefill-decode (P-D) disaggregation~\cite{PD-dist} is a standard practice in modern LLM serving for improved latency SLO attainment~\cite{PD-slo}.
Latency is reported using \textit{Time to First Token (TTFT)} for prefill instances and \textit{Time Between Tokens (TBT)} for decode instances.
Throughput is measured as input-token throughput for prefill and generated-token throughput for decode.

\begin{table}[b]
\begin{tabular}{lccc}
\toprule
\textbf{Metric} & \textbf{Mean} & \textbf{Median} & \textbf{Max} \\
\midrule
Input length (tokens)  & 13,516  & 8,001 & 12,3192 \\
Output length (tokens) & 349  & 362 & 2,000 \\
Total requests         & \multicolumn{3}{c}{3{,}000} \\
\bottomrule
\end{tabular}
\centering
\caption{Input-output characteristics of our scaled Mooncake trace.}
\label{tab:online_mooncake_stats}
\end{table}

\textbf{Dataset.} We use the real-world Mooncake~\cite{qin2025mooncake} conversation trace, a popular open-source trace with arrival timestamps of requests as well as anonymized input and output sequence lengths.
We randomly sampled 3,000 requests out of the complete traces and scale the timestamp for scanning different request rates.
Table~\ref{tab:online_mooncake_stats} summarizes the input and output characteristics of the selected trace.
Requests are issued according to the scaled timestamps, and we record TTFT, TBT, and token throughput for each system. By varying the scaling ratio, we obtain the complete throughput-latency curves.

\textbf{Baselines.}
We compare \sysname against three baselines:
(1) \textit{Standard-TP4}, the default fallback tensor-parallel configuration activated upon GPU failure, serving as the primary baseline;
(2) \textit{Standard-TP8}, the fault-free configuration representing the upper-bound performance; and
(3) \textit{Nonuniform-TP7}, a naïve implementation that operates on seven GPUs but suffers from significant load imbalance.
Our proposed \sysname adopts the optimized \textit{\sysname-TP7} configuration, which integrates all key components—cyclic memory placement, hybrid attention, and a load-aware router and scheduler.
We focus on the 7-GPU setting as it represents the most common failure scenario in practice; additional experiments under higher failure rates are presented in \S\ref{sub:less_gpu}.
Note that the \textit{Standard-TP4} baseline is omitted for the Mixtral-8x22B experiment, as the model weights and KVCache for the longest requests cannot fit within the memory capacity of only four GPUs.

\textbf{Performance Analysis.}
Figure~\ref{fig:e2e_online} presents the throughput–latency characteristics of \sysname compared with all baselines across both the prefill and decode stages.
For both the dense \textit{LLaMA-3.1-70B} and the MoE \textit{Mixtral-8x22B} models, \sysname consistently outperforms \textit{Standard-TP4} and \textit{Nonuniform-TP7} under all request rates.
At low request rates, it achieves near-optimal performance comparable to the fault-free \textit{Standard-TP8} configuration.

In the prefill stage, \sysname attains up to $2\times$ and $1.28\times$ higher throughput than \textit{Standard-TP4} and \textit{Nonuniform-TP7}, respectively, under the same 10s TTFT constraint for \textit{LLaMA-3.1-70B}, and achieves a $1.14\times$ gain over \textit{Nonuniform-TP7} for \textit{Mixtral-8x22B}.

In the decode stage, where balanced memory and compute utilization becomes more critical, the benefits are even more pronounced.
Under a 40ms TBT constraint, \sysname achieves up to $2\times$ and $1.60\times$ higher throughput than \textit{Standard-TP4} and \textit{Nonuniform-TP7} for \textit{LLaMA-3.1-70B}, and a $1.85\times$ improvement over \textit{Nonuniform-TP7} for \textit{Mixtral-8x22B}.

\subsection{Performance Breakdown Analysis}

\subsubsection{Hybrid Attention with Less GPUs}
\label{sub:less_gpu}

To further evaluate the effectiveness of our Hybrid Attention mechanism, we measure system throughput across different TP configurations.
We compare \sysname against the \textit{Nonuniform-TP} baseline under TP5–TP7 settings and report the peak throughput on the Mooncake trace using the \textit{LLaMA-3.1-70B} model.
\begin{figure}[h]
    \centering
    \includegraphics[width=\linewidth]{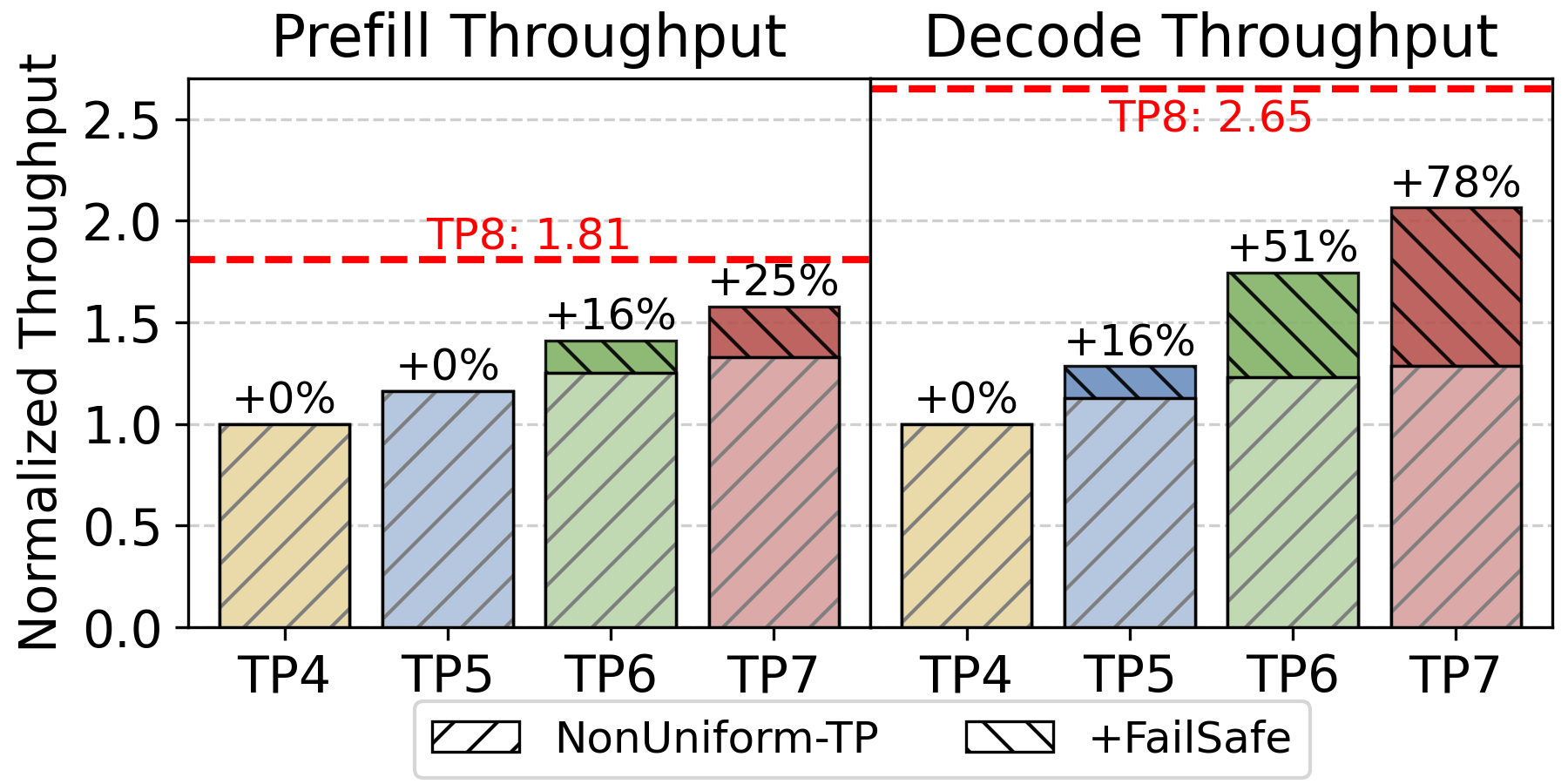}
    \vspace{-2em}
    \caption{Performance comparison of \sysname optimization with \textit{Nonuniform-TP}. Throughput is normalized to \textit{Standard-TP4}.}
    \label{fig:breakdown_throughput_tp}
\end{figure}

As shown in Figure~\ref{fig:breakdown_throughput_tp}, when running with four or eight GPUs (TP4/TP8), both systems degenerate to standard uniform tensor parallelism and therefore exhibit identical performance.
As the tensor-parallel world size becomes irregular (beyond four GPUs), computation and memory imbalance intensifies, and \sysname begins to demonstrate clear advantages.
In the prefill stage, \sysname achieves performance gains of 0\%, 16\%, and 25\% over \textit{Nonuniform-TP} for TP5, TP6, and TP7 configurations, respectively.
The limited improvement under TP5 arises because compute imbalance at this scale is inherently harder to mitigate.
In contrast, during the memory-bound decode stage, the benefits become more pronounced—\sysname improves throughput by 16\%, 51\%, and 78\% for TP5, TP6, and TP7, respectively.
These results demonstrate that the proposed hybrid attention design effectively alleviates intra-layer imbalance and scales efficiently under non-uniform GPU configurations.

\subsubsection{Memory and Compute Balancing}
Without memory balancing, certain GPUs store disproportionately larger portions of the KVCache and model weights, constraining the maximum batch size during the decode stage and leading to suboptimal GPU utilization.
To quantify the impact of memory balancing, we first augment the baseline \textit{Nonuniform-TP7} configuration with cyclic memory placement, ensuring even distribution of KVCache and model weights across GPUs.
To further highlight the effect of computation balancing introduced by our hybrid attention mechanism, we compare the peak throughput of four system configurations on the Mooncake trace using the \textit{LLaMA-3.1-70B} model:
(1) \textit{Standard-TP4} (baseline),
(2) \textit{+Nonuniform-TP7},
(3) \textit{+Memory-balancing} (\textit{Nonuniform-TP7} with cyclic memory placement), and
(4) \textit{+Compute-balancing}, which integrates all proposed optimizations.

\begin{figure}[h]
    \centering
    \includegraphics[width=\linewidth]{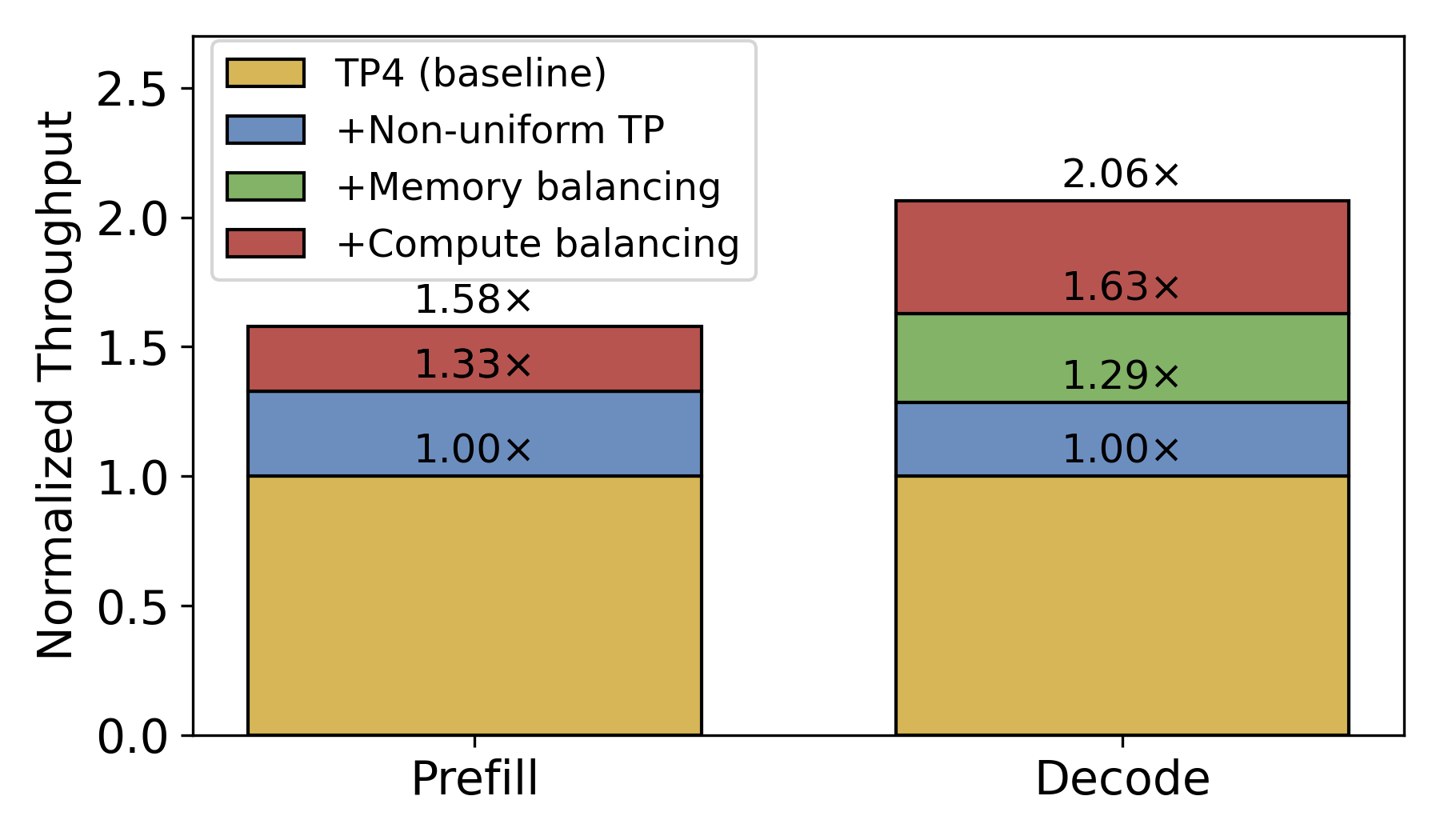}
    \vspace{-2em}
    \caption{Breakdown of \sysname optimization attributions in both prefill stage and decode stage on Llama-70B model.}
    \label{fig:breakdown_throughput_bar}
\end{figure}

As shown in Figure~\ref{fig:breakdown_throughput_bar}, memory and compute balancing contribute differently across stages.
In the prefill stage, the compute-balancing mechanism alleviates straggler effects and improves peak throughput by 25\%, while memory balancing provides negligible benefit since prefill is primarily compute-bound.
In contrast, the decode stage is memory-bound: larger batch sizes enabled by memory balancing significantly improve GPU utilization, yielding a 34\% throughput increase.
Finally, applying compute balancing on top of memory balancing further mitigates inter-GPU workload skew, boosting overall decode throughput by an additional 43\%.

\subsubsection{Recovery Latency}
\label{sec:recovery_latency}
To evaluate the efficiency of our recovery mechanism, we conduct a detailed breakdown of recovery latency under online serving workloads, where latency sensitivity is most critical.
We use the \textit{LLaMA-3.1-70B} model and replay a contiguous 500-request window from the Mooncake conversation trace.
The system initially runs with a standard TP8 decode instance, and a GPU failure is injected 100ms after the 250th request (i.e., halfway through the trace).
We report the maximum \textit{Time Between Tokens (TBT)} per request as the latency metric, since a request is considered to violate its decode SLO if any of its TBTs exceed the specified threshold.

All configurations enable memory and compute balancing optimizations.
We compare the following recovery methods:
(1) \textit{Recompute}, which regenerates the lost KVCache and reloads all re-sharded model weights when transitioning to a TP7 configuration;
(2) \textit{\sysname-Host}, which proactively backs up the KVCache in host memory and restores the lost cache from backup instead of recomputation;
(3) \textit{\sysname-Full}, which builds upon \textit{\sysname-Host} and further avoids redundant PCIe transfers through joint on-demand weight loading; and
(4) \textit{\sysname-Oracle}, which represents an idealized setting that restores only necessary metadata to the GPU without performing any weight or KVCache loading.

\begin{table}[ht]
\centering
\caption{GPU state recovery latency of different systems.}
\label{tab:recovery}
\begin{tabular}{lcccc}
\toprule
\textbf{System} & \textit{Recompute} & \textit{Host} & \textit{Full} & \textit{Oracle} \\ 
\midrule
\textbf{Latency} & 22~s & 530~ms & 120~ms & 15~ms \\
\textbf{Speedup} & 1.00$\times$ & 41.5$\times$ & 183$\times$ & N/A \\
\bottomrule
\end{tabular}
\end{table}

\begin{figure}[h]
    \centering
    \includegraphics[width=\linewidth]{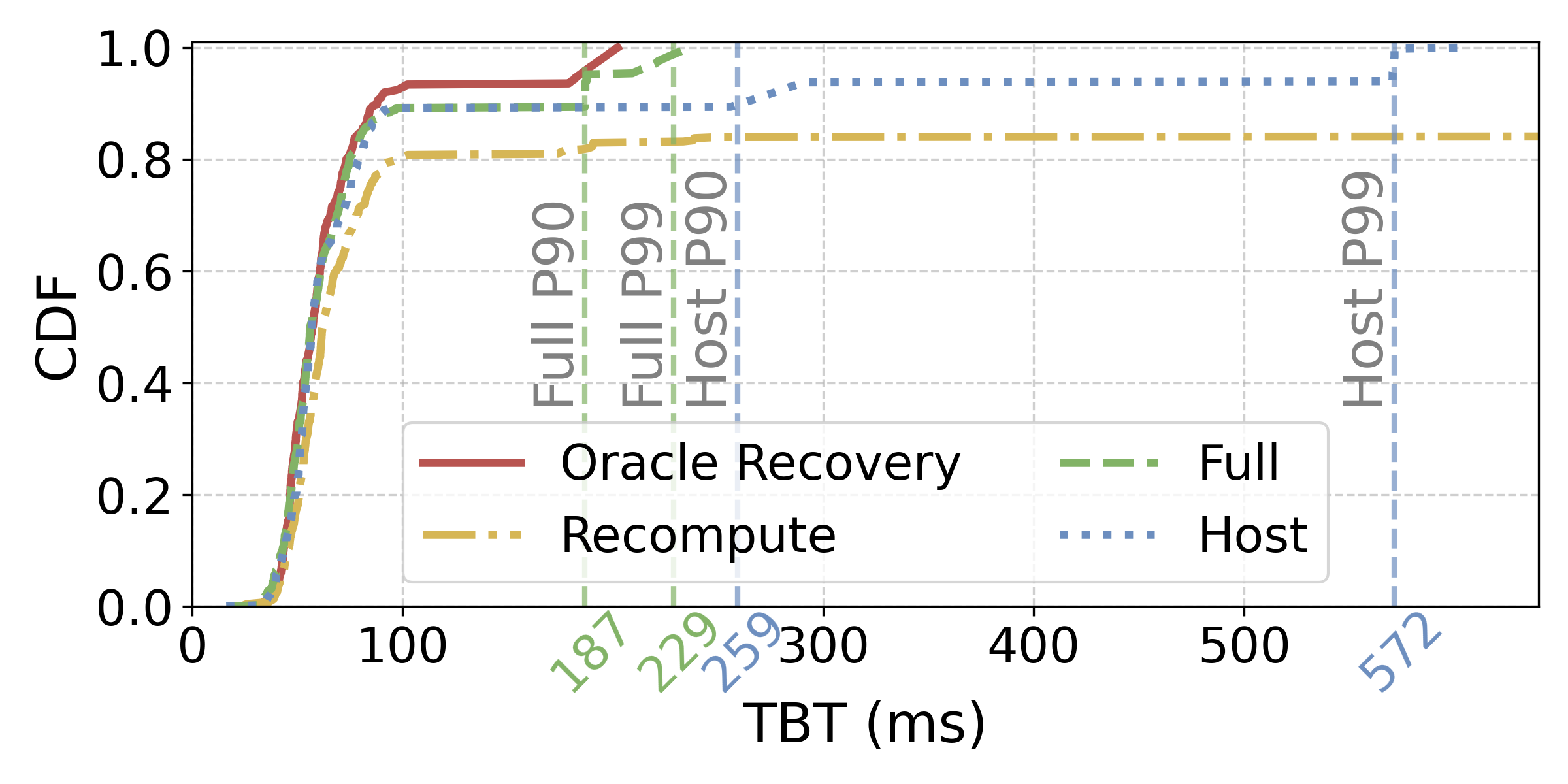}
    \vspace{-2em}
    \caption{CDF of Max TBT per request of \sysname under different recovery method. Dashed vertical lines shows the P90/P99 TBT of \textit{\sysname Full} and \textit{\sysname Recompute}.
    }
    \label{fig:breakdown_recovery_cdf}
\end{figure}

Table~\ref{tab:recovery} summarizes GPU recovery latencies across settings.
Compared to \textit{Recompute}, restoring backed-up KVCache from host memory is $41.5\times$ faster, while \textit{\sysname-Full} further reduces recovery latency by an additional $4.4\times$.
These improvements translate directly into user-perceived latency reductions.
As shown in Figure~\ref{fig:breakdown_recovery_cdf}, proactive host-side KVCache backup eliminates expensive recomputation, reducing P90 and P99 TBT from over 10s to under 1s.
With on-demand weight loading, P99 TBT further drops from 572ms to 229ms, a 2.5$\times$ improvement and approaching the oracle’s lower bound.
Together, these results confirm that host-based KVCache backup and on-demand recovery are highly effective in minimizing recovery latency during online serving.

\section{Related Work}

Fault tolerance in distributed training has been extensively studied by numerous systems. Redundancy-based methods such as Bamboo~\cite{bamboo} incorporate redundant computations to handle preemptions common in cloud spot instances. Oobleck~\cite{oobleck} employs heterogeneous pipelines to facilitate failure recovery without requiring additional spare resources, while ReCycle~\cite{recycle} dynamically reroutes computations to redundant data-parallel peers to seamlessly continue processing after failures. Checkpoint-based approaches include CheckFreq~\cite{checkfreq}, CPR~\cite{cpr}, and Gemini~\cite{gemini}, which respectively mitigate checkpoint overhead by dynamically adjusting checkpoint frequency, quantizing embedding tables, and strategically scheduling checkpoint traffic across the storage hierarchy. Megascale~\cite{megascale} addresses storage bottlenecks during recovery by efficiently sharing data between corresponding GPU workers across data-parallel groups. Unlike these training-focused systems, \sysname specifically employs hybrid attention mechanism with non-uniform tensor parallelism to achieve fault-tolerant distributed inference.

In the context of fault-tolerant distributed inference, recent efforts such as SpotServe~\cite{spotserve}, Llumnix~\cite{llumnix}, DejaVu~\cite{dejavu}, and Medusa~\cite{medusa} enable live migration of KVCaches and model weights. However, these systems do not address the persistent throughput degradation arising from compute and memory imbalances inherent in non-uniform tensor parallelism. To overcome these challenges, \sysname integrates a hybrid attention mechanism and a dynamic scheduler alongside KVCache migration, ensuring both high-performance and robust resilience for distributed inference workloads.

\section{Discussion}
\label{sec:discussion}

Having demonstrated the advantages of \sysname, we now discuss its applicability, limitations, and directions for future work.
While tensor parallelism remains the dominant approach for scaling large models across multiple GPUs, it is not the only viable strategy. Other parallelism techniques offer inherent fault tolerance and may better suit specific model architectures.
For instance, recent studies show that expert parallelism can achieve performance comparable to TP~\cite{ep} for large MoE models, while exhibiting stronger resilience to partial GPU loss.
Although our focus has been on fault-tolerant serving, \sysname can also facilitate finer-grained resource sharing among concurrent jobs, mitigating starvation caused by rigid gang scheduling~\cite{jeon2018multi}.
Integrating \sysname with dynamic GPU resource allocation represents a promising direction for future exploration.
As hardware platforms continue to scale, we expect \sysname to become increasingly relevant for large multi-GPU systems such as NVL72 \cite{NVL72}.
While our evaluation focuses on single-node configurations, extending memory and compute balancing beyond NUMA boundaries remains an important avenue for future work.

\section{Conclusion}

We present \sysname, a high-performance yet resilient serving system for tensor-parallel LLM inference under irregular GPU availability. \sysname tackles the recovery overhead by introducing proactive KVCache backup and on-demand weight recovery, accelerating the recovery process by 183$\times$ and thus avoiding a devastating latency spike in online serving.
\sysname also introduces cyclic KVCache placement, hybrid attention, and fine-grained load-aware router to eliminate the memory and computation imbalance between GPUs, achieving higher memory and computation utilization and up to
2× higher throughput and two orders of magnitude faster recovery than standard fault-handling methods.
Looking ahead, the principles behind \sysname extend naturally to larger heterogeneous environments and to emerging architectures such as NVL72.

%%
%% Bibliography
%%
\bibliography{references}
\bibliographystyle{mlsys2025}

\end{document}